\documentclass{PoS}

\def\l{{\boldsymbol l}}

\def\pT{p_{_T}}

\def\Raa{Q_\mathrm{AA}}

\newcommand{\beq}{\begin{eqnarray}}
\newcommand{\eeq}{\end{eqnarray}}
\newcommand{\be}{\begin{eqnarray*}}
\newcommand{\ee}{\end{eqnarray*}}
\newcommand{\bal}{\begin{align}}
\newcommand{\eal}{\end{align}}

\newcommand{\dd}{{\rm d}}

\makeatletter
\def\tagform@#1{\maketag@@@{\ignorespaces#1\unskip\@@italiccorr}}

\makeatother

\title{Jet quenching and scaling properties of medium-evolved gluon
cascade in expanding media}

\ShortTitle{Jet quenching and scaling properties of medium-evolved gluon
cascade in expanding media}

\author{\speaker{Souvik Priyam Adhya}\\
        Institute of Particle and Nuclear Physics, Faculty of Mathematics and Physics, Charles University, V Holesovickach 2, Prague, 18000, Czech Republic\\
        E-mail: \email{souvik@ipnp.mff.cuni.cz}}
\author{Carlos A. Salgado\\
        Departamento de Fisica de Particulas and IGFAE, Universidade de Santiago de Compostela, 15782 Santiago de Compostela, Spain\\}
        \author{Martin Spousta\\
         Institute of Particle and Nuclear Physics, Faculty of Mathematics and Physics, Charles University, V Holesovickach 2, Prague, 18000, Czech Republic\\}
\author{Konrad Tywoniuk\\
        Department of Physics and Technology, University of Bergen, Postboks 7803, 5020 Bergen, Norway\\}

\abstract{We present a study of the impact of the expansion of deconfined medium on single-gluon emission spectra and the jet suppression factor ($Q_{AA}$) within the BDMPS-Z formalism. These quantities are calculated for three types of media (static medium, exponentially decaying medium and Bjorken
expanding medium). The distribution of medium-induced gluons and the jet $Q_{AA}$ are calculated using the evaluation of in-medium evolution with splitting kernels derived from the gluon emission spectra. Scaling behavior of splitting kernels is derived for low-x and high-x regimes in the asymptote of large times and its impact on the resulting jet $Q_{AA}$ is discussed. For the full phase space of the radiation, the scaling of jet $Q_{AA}$ with an effective quenching parameter is presented.}

\FullConference{10th International Conference on Hard and Electromagnetic Probes of High-Energy Nuclear Collisions\\
		1-6 June 2020\\
		Austin, Texas}

\begin{document}

Newer theoretical tools for understanding the medium induced radiative energy loss from the perspective of jet quenching in RHIC and LHC have been studied in detail in recent years \cite{Baier:1996sk,Zakharov:1996fv,Salgado:2003gb,Wiedemann:2000za, Mehtar-Tani:2014yea, Arnold:2008iy, Blaizot:2013hx,Andres:2020vxs, Adhya:2019qse}. In light of these recent developments of  characterizing the soft and hard processes of partonic energy loss in media, we calculate the medium modified gluon spectra with splitting kernels derived analytically for dynamically expanding media. Next, we present results discussing the validity of the scaling parameters, derived previously in \cite{Salgado:2003gb,Salgado:2002cd} for the full kinematical spectra. Finally, we present calculations of the jet suppression factor and its scaling for the expanding media. In depth details about this work can be found in \cite{Adhya:2019qse}. We restrict our calculations to gluon splittings and thus results of $\Raa$ presented here serve as a proxy for jet $R_{AA}$ as in experiments. The conclusions regarding the scaling behavior among different profiles point to the importance of the role of medium expansion in characterizing the jet quenching phenomena.
%

The time dependency of the jet quenching parameter for the static and exponentially expanding media can be written as,
\begin{eqnarray}
&\hat q(t)^{static} = \hat q_0,\\
&\hat q(t)^{expo} = \hat q_0 \exp{(-t/L)}.
\end{eqnarray}
and the Bjorken expanding medium as,
\begin{eqnarray}
&\hat q(t)^{Bjorken} = 0 \hspace{2.5cm}& {\rm for } \quad t<t_0 \,, \\ 
& \hspace{0.5cm}=\hat q_0 (t_0/t)^\alpha & {\rm for} \quad t_0 < t < L+t_0 \,,\nonumber \\
&\hspace{-0.5cm}= 0 & {\rm for} \quad L+t_0 < t \,.\nonumber
\end{eqnarray}
where $\hat q_0$ is the quenching parameter for the static media and $L$ is the length of the media traversed by the initial parton. 
For our purpose, we introduce the single gluon emission spectra for different profiles of the media as follows \cite{Arnold:2008iy,Adhya:2019msi,Adhya:2019qse}
\begin{eqnarray}
\frac{\dd I}{\dd z}^{static} &=& \frac{\alpha_s}{\pi} P(z)\, \mathrm{Re} \ln \cos \Omega_0 L \,, \\
\frac{\dd I}{\dd z}^{exponential} &=& \frac{\alpha_s}{\pi} P(z)\, \mathrm{Re} \ln  J_0( 2\Omega_0 L ) \,, \\
\frac{\dd I}{\dd z}^{Bjorken} &=& \frac{2\alpha_s}{\pi} P(z)\, \mathrm{Re} \ln\left[ \left(\frac{t_0}{L+t_0} \right)^{1/2} \frac{J_1(z_0)Y_0(z_L) - Y_1(z_0) J_0(z_L)}{J_1(z_L)Y_0(z_L) - Y_1(z_L) J_0(z_L)} \right]\,
\end{eqnarray}
where $P(z)\equiv P_{gg}(z)$ is the Altarelli-Parisi function for the gluon- gluon splitting.
We can also derive an emission rate for the gluons, defined as
\beq 
\mathcal{K}(z,\tau) \equiv \frac{\dd I}{\dd z \dd \tau} \,, 
\eeq 
where the evolution parameter $\tau$ ($\tau = \sqrt{\hat q_0/p}L $) is a dimensionless variable.
Therefore, we can derive analytically different splitting kernels as,
\begin{eqnarray}
 \mathcal{K}(z,\tau)^{static} &=& \frac{\alpha_s}{2\pi} P(z) \kappa(z)\, \mathrm{Re} \left[(i-1) \tan \big((1-i)\kappa(z) \tau/2 \big) \right]\,.\\
 \mathcal{K}(z,\tau)^{exponential} &=& \frac{\alpha_s}{\pi} P(z) \kappa(z) \,\mathrm{Re} \left[ (i-1) \frac{J_1\big((1-i)\kappa(z) \tau \big)}{J_0\big((1-i)\kappa(z) \tau \big)} \right]\\
\mathcal{K}(z,\tau)^{Bjorken} &=& \frac{\alpha_s}{\pi} P(z) \kappa(z)\sqrt{\frac{\tau_0}{\tau + \tau_0}} \nonumber\\
&&\times \mathrm{Re}\left[ (1-i) \frac{J_1(z_L) Y_1(z_0) - J_1(z_0) Y_1(z_L) }{J_1(z_0) Y_0(z_L) - J_0(z_L) Y_1(z_0)} \right]\\
\end{eqnarray}

Here, we re-define the evolution time $\tau$ as $\tau_{eff}$ with the relation  $2\tau$ and $2\sqrt{\tau_0\tau}$ for the exponential and Bjorken profiles respectively. We will use "soft scaling" to refer to scalings introduced by this definition of evolution time for expanding media profiles in the rest of the paper. 
\begin{figure}
\centering
\includegraphics[width=7.2cm,height=5.2cm]{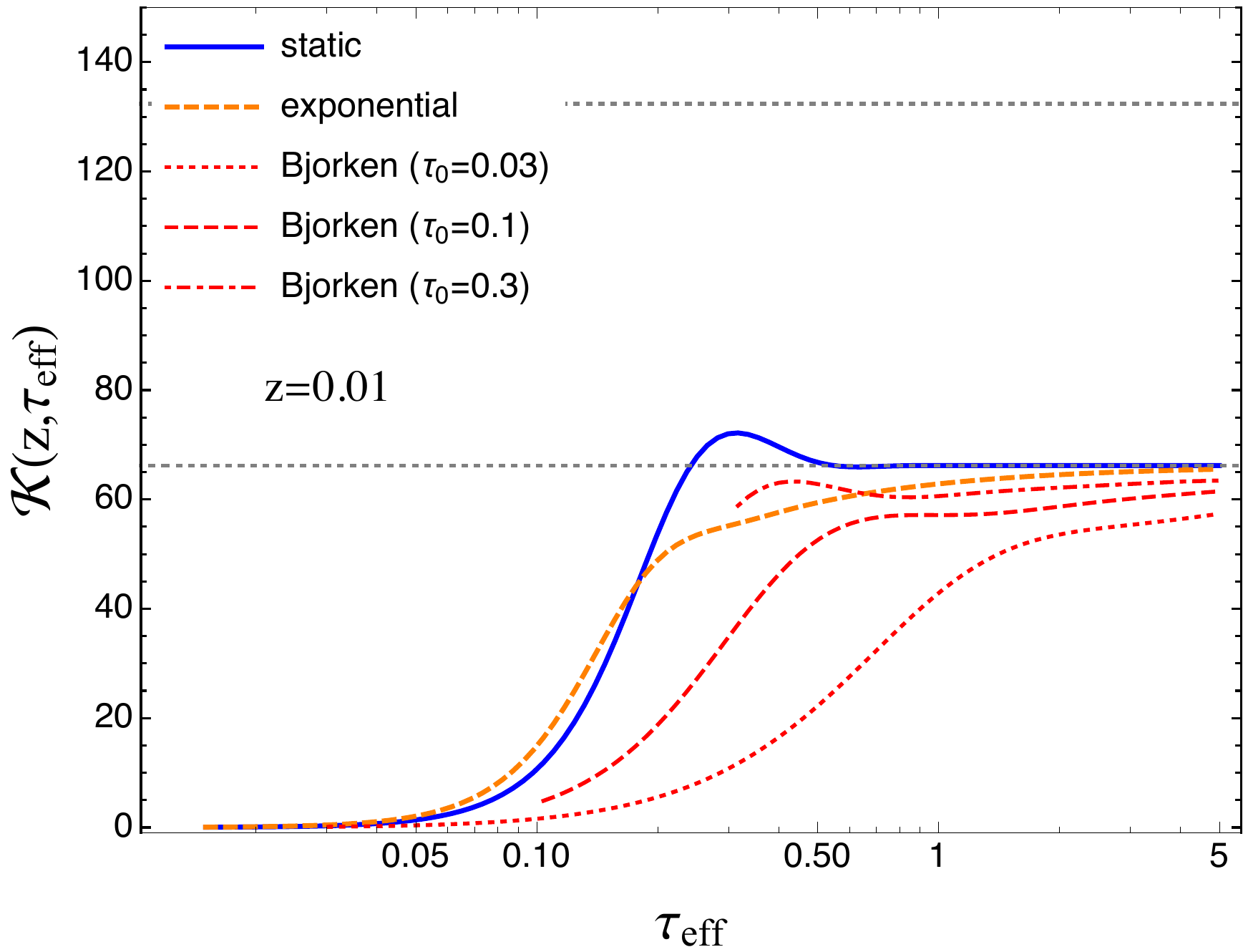}~
\includegraphics[width=7.2cm,height=5.2cm]{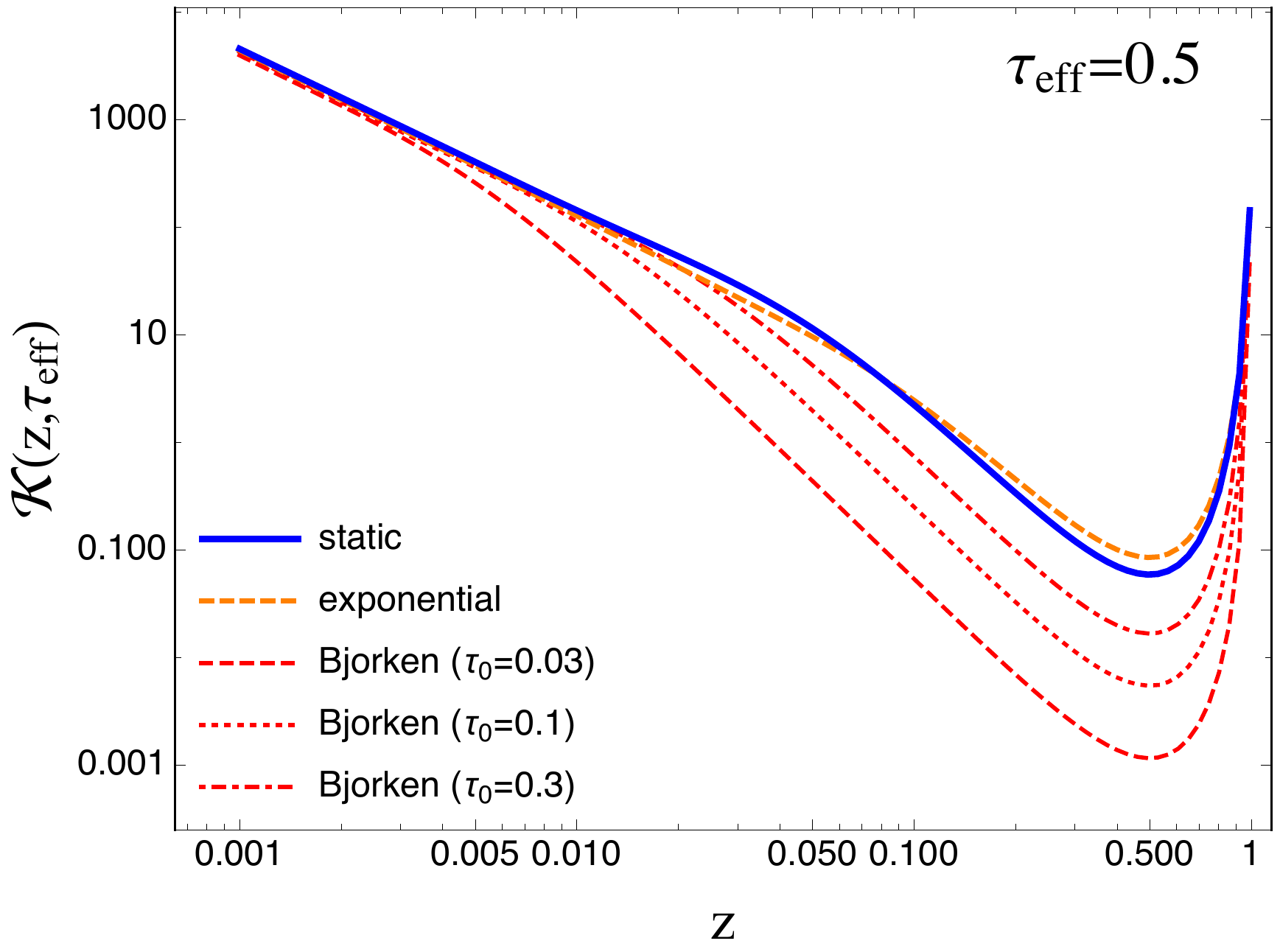}
\caption{Scaling features of the gluon emission rate for effective evolution time $\tau_{eff}=0.5$ (left) and  momentum fraction $z$ (right).}
\label{fig:rate-z}
\end{figure}

In Fig.(1), we plot the splitting rates as a function of $\tau_{eff}$ and momentum fraction $z$ respectively. We observe  a universal scaling feature only in the soft gluon regime for all the profiles. For a comprehensive discussion on the scaling aspects of the spectra and rates, refer to \cite{Adhya:2019qse}. 

Next, resumming multiple  gluon emissions within the media via the kinetic rate equation given by \cite{Blaizot:2013hx},
\begin{equation}
\label{eq:RateEquation-generic}
\frac{\partial D(x, \tau)}{\partial \tau} = \int_0^1 \dd z \,\mathcal{ K}(z,\tau) \left[\sqrt{\frac{z}{x}} D\left(\frac{x}{z},\tau \right) \Theta(z-x) - \frac{z}{\sqrt{x}} D(x,\tau) \right] \,.
\end{equation}
we numerically evaluate the medium modified gluon distribution as presented in Fig.(2). The initial condition for $D(x,\tau)$ is set to a $\delta$-function at $x=1$. An effective scaling is possible for the medium modified gluon distribution for the singular kernels. However, the scaling does not hold for the distribution plotted with the full kernels. For a detailed explanation, see \cite{Adhya:2019qse}.
\begin{figure}
\centering
\includegraphics[width=7.2cm,height=5.2cm]{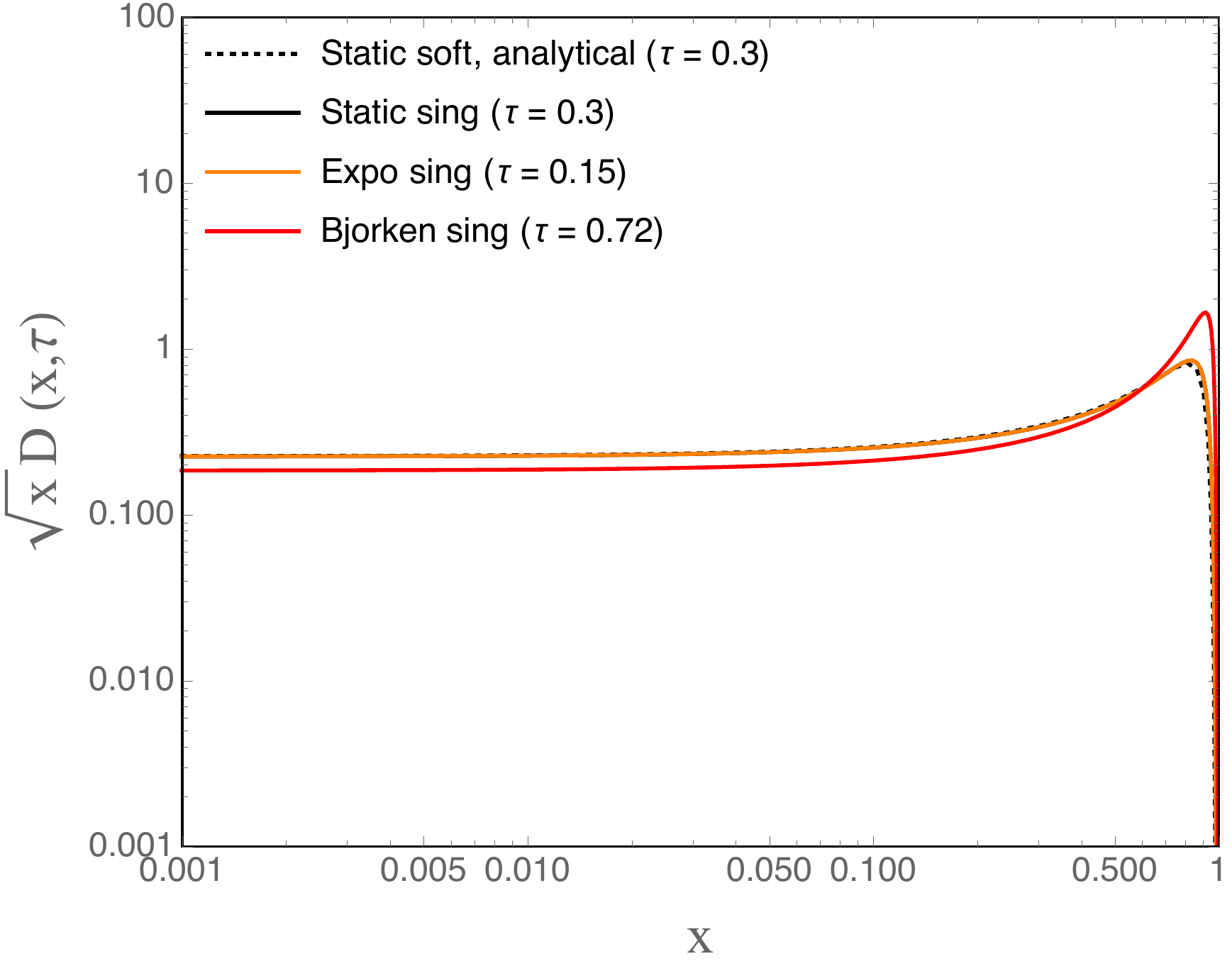}~
\includegraphics[width=7.2cm,height=5.2cm]{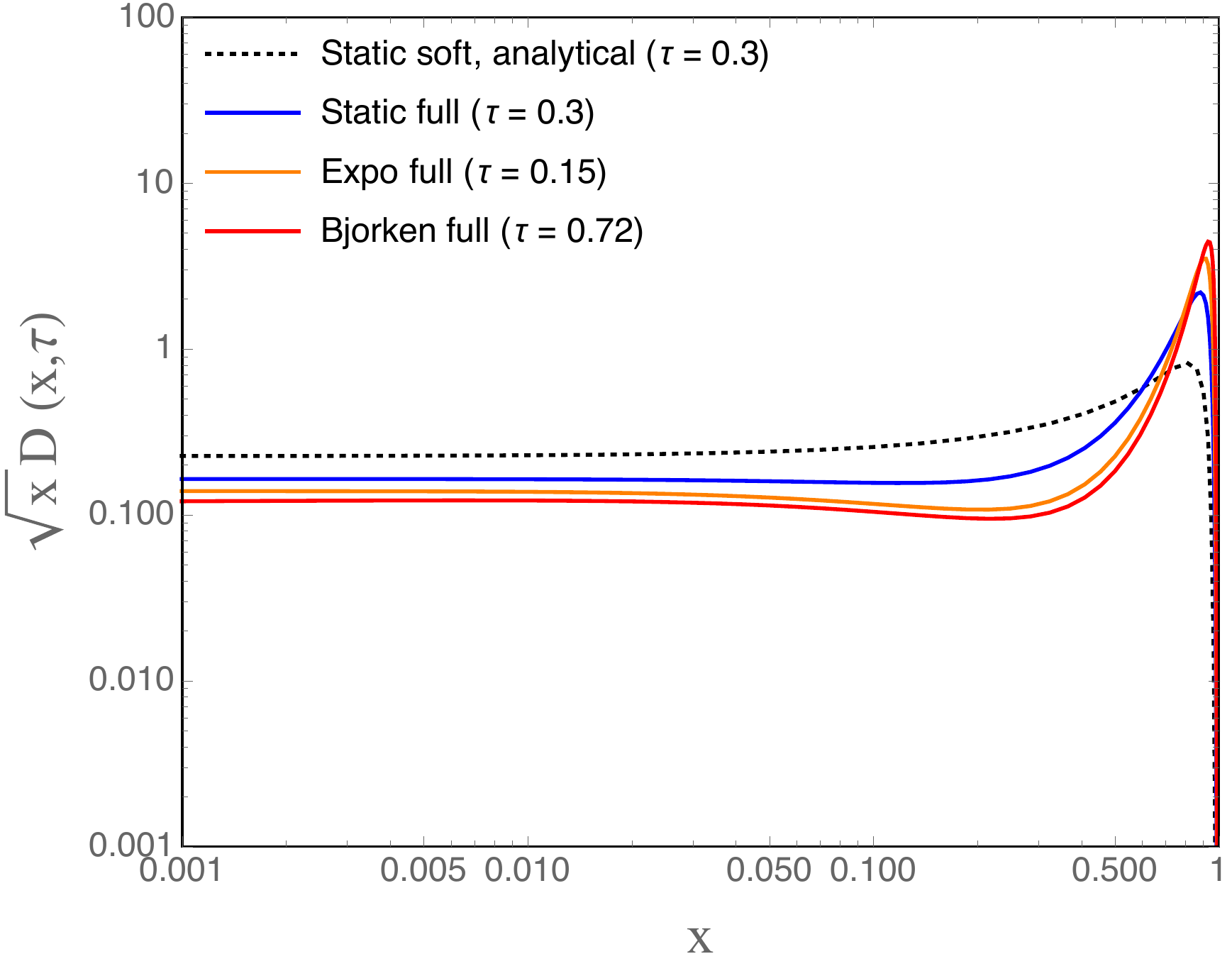}
\caption{
Medium modified gluon spectra $\sqrt{x}D(x,\tau)$ for singular rates (left panel) and full rates (right panel) for different evolution time $\tau$ (with  appropriate scaling for expanding media profiles).}
\label{fig:distributions-scaling}
\end{figure}
Finally, we use the medium modified gluon distribution for calculating the quenching factor for the jets given as,
\beq
\label{eq:suppression-factor-1}
\Raa(\pT) = \int_0^1 \dd x \, x^{n-1} D(x, \sqrt{x} \tau) \,
\eeq
\begin{figure}[tbp]
\centering
\includegraphics[width=5.2cm,height=5.2cm]{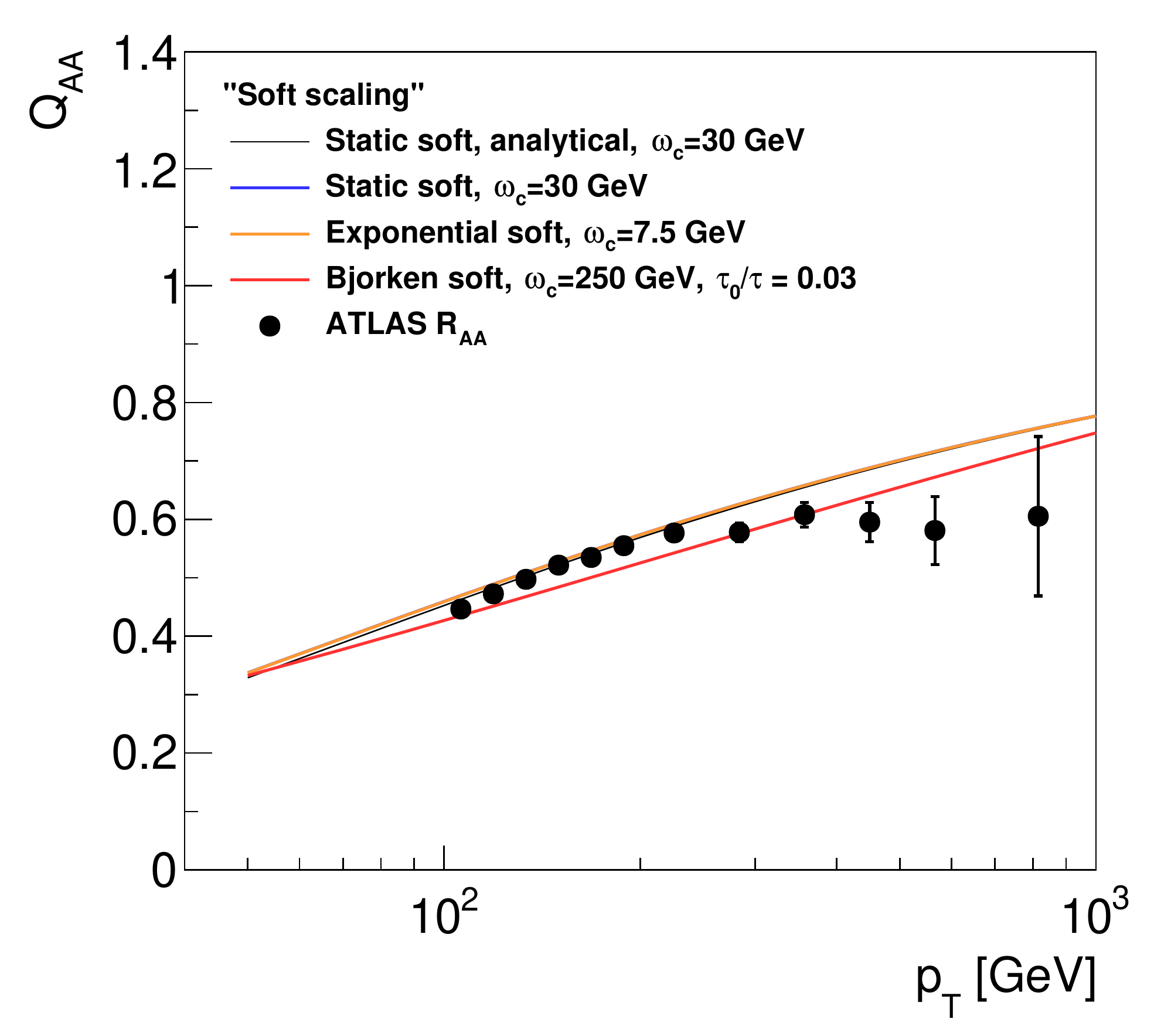}~
\includegraphics[width=5.2cm,height=5.2cm]{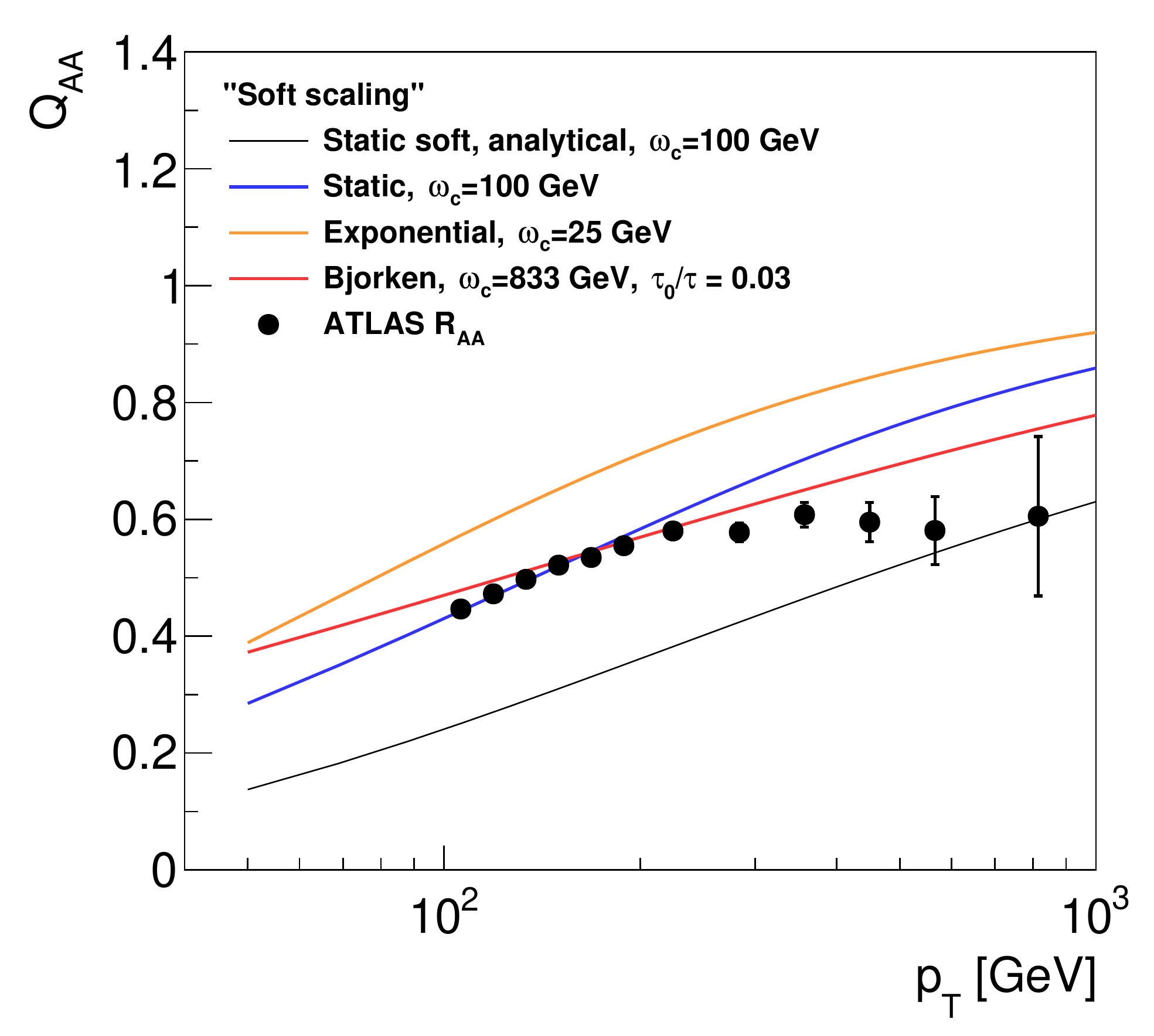}~
\includegraphics[width=5.2cm,height=5.2cm]{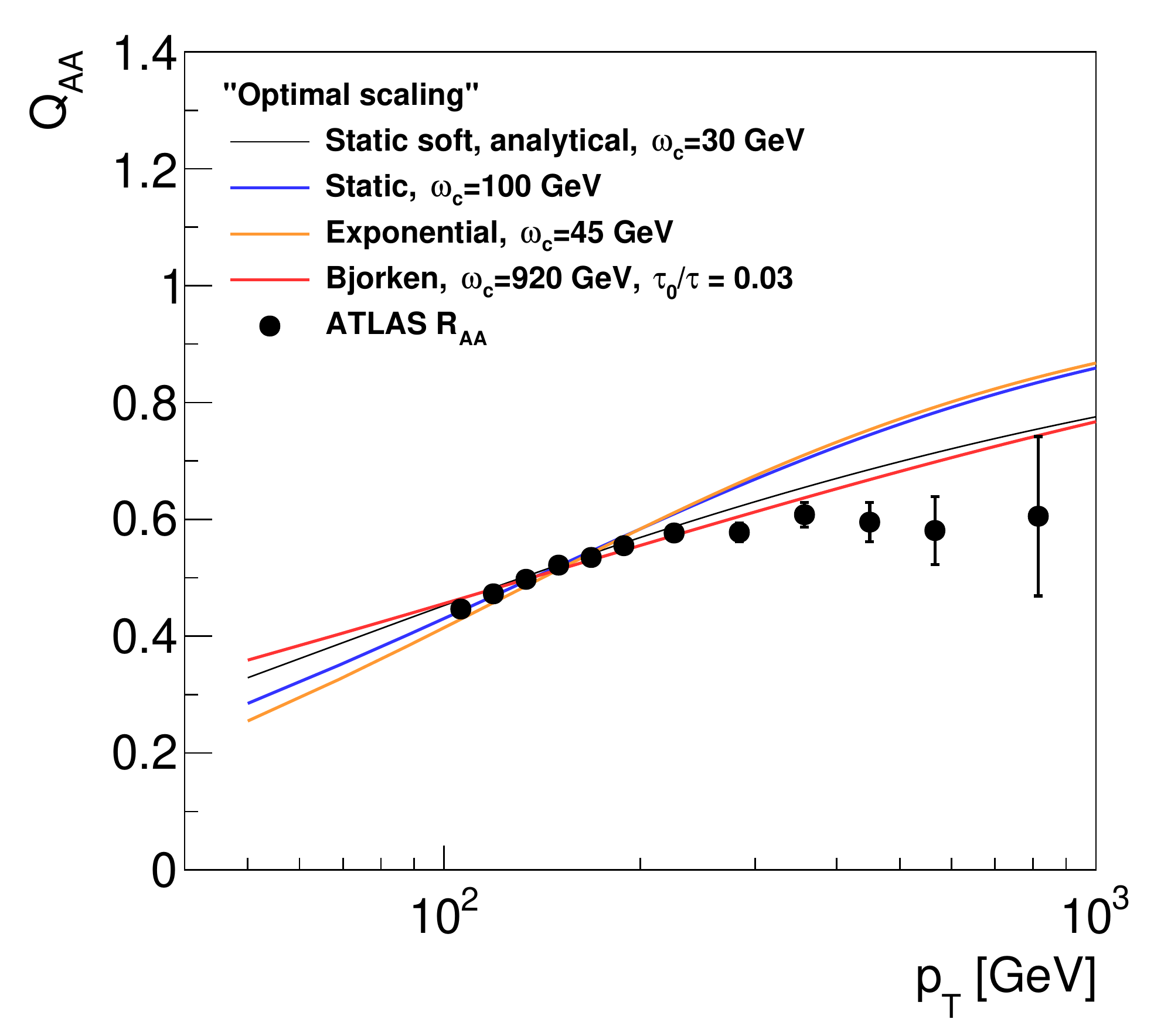}
\caption{The $\Raa$ distributions using singular splitting kernels (left), full splitting kernels with "soft scalings" (middle) and full kernels with "optimal scaling" (right) for all the medium profiles. The plots also include a comparison with the ATLAS data \cite{Aaboud:2018twu}.}
\label{fig:raa2}
\end{figure}
In the left panel of Fig.(3), we present plots where we implemented the "soft scaling" (with $\tau_{eff}$) in the singular kernel for the different medium expansions. In the middle panel, we use the same effective scaling parameters in determining the suppression factor with the full splitting kernels. Finally, in the right panel, we introduce an "optimal scaling" to allow minimal differences among different profiles in $\Raa$. We note that for the exponential case, the "average scaling" (derived from average value of $\hat{q}$) holds better than "soft scaling" for full kernels \cite{Adhya:2019qse}. We conclude that for the Bjorken case, which is quite sensitive to the choice of the initial parameter $\tau_0$, no scaling holds perfectly over the full kinematical range.

\section*{Acknowledgement} 
 KT is supported by a Starting Grant from Trond Mohn Foundation 
(BFS2018REK01) and the University of Bergen. CAS is supported by Ministerio de Ciencia e Innovaci\'on of Spain under project FPA2017-83814-P; Unidad de Excelencia Mar\'ia de Maetzu under project MDM-2016-0692;
ERC-2018-ADG-835105 YoctoLHC; and Xunta de Galicia (Conseller\'ia de Educaci\'on) and FEDER. SPA and MS are supported by Grant Agency of the Czech Republic under grant 18-12859Y, by the Ministry of Education, Youth and Sports of the Czech Republic under grant LTT~17018, and by Charles University grant UNCE/SCI/013.

\end{document}